\pdfoutput=1
\documentclass[twocolumn, pra, pdflatex]{revtex4}
\usepackage{graphicx}
\usepackage{amsmath,amssymb}

\begin{document}

\title{Temporal coherence and indistinguishability in two-photon interference effects}
\author{Anand Kumar Jha, Malcolm N. O'Sullivan, Kam Wai Clifford Chan, and Robert W. Boyd}
\affiliation{The Institute of Optics, University of Rochester,
Rochester, New York 14627, USA}
\date{\today}

\begin{abstract}

We show that temporal two-photon interference effects involving
the signal and idler photons created by parametric down-conversion
can be fully characterized in terms of the variations of two
length parameters---called the biphoton path-length difference and
the biphoton path-asymmetry-length difference---which we construct
using the six different length parameters that a general
two-photon interference experiment involves. We perform an
experiment in which the effects of the variations of these two
parameters can be independently controlled and studied. In our
experimental setup, which does not involve mixing of signal and
idler photons at a beam splitter, we further report observations
of Hong-Ou-Mandel- (HOM-)like effects both in coincidence and in
one-photon count rates. As an important consequence, we argue that
the HOM and the HOM-like effects are best described as
observations of how two-photon coherence changes as a function of
the biphoton path-asymmetry-length difference.

\end{abstract}


\maketitle

In the past few decades, much attention has been devoted to the
study of two-photon interference effects involving the
signal-idler pair of photons (also called biphotons) produced by
parametric down-conversion (PDC) \cite{quantum effects, HOM,
franson interferometer, kwiat-interference, Martienssen, induced
coherence, frustrated two-photon, shih-can two-photon, shih-dip,
shih-single-photon, geometric}. The Hong-Ou-Mandel (HOM) effect
\cite{HOM}, two-photon fringes in a Franson interferometer
\cite{franson interferometer, kwiat-interference, Martienssen},
induced coherence without induced emission \cite{induced
coherence}, frustrated two-photon creation \cite{frustrated
two-photon}, and postponed compensation \cite{shih-can two-photon}
are some of the very interesting temporal two-photon interference
effects observed among many others. All of these effects have been
explained using the mathematical framework worked out by Glauber
\cite{glauber1}. However, different conceptual pictures have often
been used to describe them.

In this Rapid Communication, we construct two length
parameters---the biphoton path-length difference and the biphoton
path-asymmetry-length difference---in terms of the six different
length parameters that a general two-photon interference
experiment involves. We show that temporal two-photon interference
effects, independent of the experimental setup, can be fully
characterized in terms of the variations of these two parameters
only. We perform an experiment in a double-pass setup in which the
variations of these two parameters can be independently controlled
and studied. In this setup, which does not involve mixing of
signal and idler photons at a beam splitter, we further report
experimental observations of ``HOM-like'' effects. HOM \cite{HOM}
and HOM-like effects \cite{shih-can two-photon,shih-dip} have so
far been observed only in those setups that involve mixing of
signal and idler photons, arriving either simultaneously
\cite{HOM} or in a postponed manner \cite{shih-can two-photon}, at
a beam splitter. As an important consequence, we argue that HOM
and HOM-like effects can be best understood as observations of how
the degree of two-photon coherence changes as a function of the
biphoton path-asymmetry-length difference defined below.
\begin{figure}[b]
\includegraphics{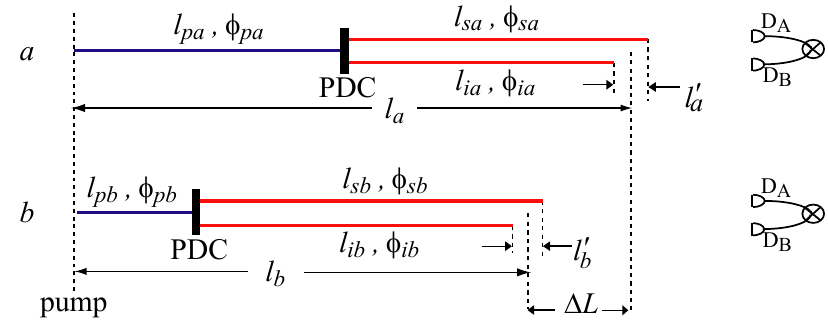}
  \caption{ (Color online) Schematic representation of two-photon
  interference using two-photon path diagrams. Alternatives \textit{a} and
  \textit{b} are the two pathways by which a pump
  photon is down-converted and the down-converted photons are
  detected at single-photon detectors
$D_{\textrm{A}}$ and $D_{\textrm{B}}$ in coincidence.
}\label{two-photon-path-diagram}
\end{figure}

We begin by describing two-photon interference in terms of the
superposition of two-photon probability amplitudes
\cite{glauber1,shih-cw} and represent a general two-photon
two-path interference experiment by the two-photon path diagrams
of Fig.~\ref{two-photon-path-diagram}. Diagrammatic approaches
have previously also been used to describe two-photon interference
effects (see Refs. \cite{shih-can two-photon,shih-dip}). For
conceptual clarity, we consider only the polarization-independent,
temporal two-photon interference effects, assuming perfect spatial
coherence. Alternatives $a$ and $b$ are the two pathways by which
a pump photon gets down-converted and the down-converted signal
and idler photons get detected at single-photon detectors
$D_{\textrm{A}}$ and $D_{\textrm{B}}$. Two-photon interference is
observed in the coincidence count rate of detectors
$D_{\textrm{A}}$ and $D_{\textrm{B}}$ as long as these two
alternatives are coherent, i.e., indistinguishable from each
other. In a two-photon interference experiment, these alternative
pathways can be introduced by using beam splitters
\cite{HOM,shih-can two-photon}, by passing the pump beam twice
through a crystal \cite{frustrated two-photon}, or even by using
two different crystals \cite{induced coherence}. We take the pump
wave to have a Gaussian spectrum with a coherence length
$l_{\textrm{coh}}^{p}$. We also assume that the spectral width of
the pump field is much smaller than that of the signal and idler
fields. In Fig.~\ref{two-photon-path-diagram}, the subscripts $p$,
$s$ and $i$ stand for pump, signal, and idler, respectively; $l$
denotes the optical path length traveled by a photon; and $\phi$
stands for phases other than the dynamical one such as phase
acquired due to reflections, geometric phase
\cite{geometric,geometric-interference}, etc. Thus $l_{sa}$
denotes the optical path length traveled by the signal photon in
alternative $a$, etc. The various path lengths and phases can be
used to define two length parameters and one phase parameter as
follows:
\begin{eqnarray}\label{define}
\Delta L  &\equiv&  l_{a}-l_{b}= \left(\frac{l_{sa}+l_{ia}}{2}+l_{pa}\right)-\left(\frac{l_{sb}+l_{ib}}{2}+l_{pb}\right), \nonumber  \\
\Delta L' &\equiv&
l'_{a}-l'_{b}=\left(l_{sa}-l_{ia})-(l_{sb}-l_{ib}\right), \nonumber\\
\Delta\phi &\equiv&
\left(\phi_{sa}+\phi_{ia}+\phi_{pa}\right)-\left(\phi_{sb}+\phi_{ib}+\phi_{pb}\right).
\end{eqnarray}
\noindent Here $l_{a(b)}$ and $l'_{a(b)}$ are the biphoton path
length and the biphoton path-asymmetry length for alternative $a
(b)$. In a particular alternative, the biphoton path length is
defined to be the mean of the optical path lengths traveled by the
signal and idler photons added to the optical path length traveled
by the pump photon. The biphoton path-asymmetry length is defined
to be the difference of the optical path lengths traveled by the
signal and idler photons. $\Delta L$ is the difference of the
biphoton path lengths $l_{a}$ and $l_{b}$ whereas $\Delta L'$ is
the difference of the biphoton path-asymmetry lengths $l'_{a}$ and
$l'_{b}$. Notice that, if either $\Delta L$ or $\Delta L'$ is too
large, alternatives $a$ and $b$ will become distinguishable and
will no longer interfere.

To quantify this point, we calculate the coincidence count rate
$R_{\textrm{AB}}$ for detectors $D_{\textrm{A}}$ and
$D_{\textrm{B}}$ in terms of $\Delta L$ and $\Delta L'$. The
detailed calculations will be presented elsewhere. However, it can
be shown by generalizing the calculations of Refs. \cite{vacuum
effects,induced coherence theory} that
\begin{eqnarray}\label{coin-count-rate}
 R_{\textrm{AB}} = C [ 1+ \gamma(\Delta L)\gamma '(\Delta
L')
  \cos({k_{0}\Delta L+k_{d}\Delta L'+\Delta \phi})].
\end{eqnarray}
Here \textit{C} is a constant, $k_{0}$ is the mean vacuum
wave-vector magnitude of the pump wave while $k_{d}\equiv
(k_{s0}-k_{i0})/2$, where $k_{s0}$ and $k_{i0}$ are the mean
vacuum wave-vector magnitudes of the signal and idler fields.
$\gamma (\Delta L)$ and $\gamma' (\Delta L')$ are the normalized
correlation functions of the pump and the signal-idler fields,
respectively. The product $\gamma' (\Delta L')\gamma (\Delta L)$
is the degree of two-photon coherence. In the rest of this Rapid
Communication, we consider only degenerate PDC in which case
$k_{d}=0$. Also, for simplicity, we consider only type I phase
matching and assume that the signal-idler field has a Gaussian
spectrum with a coherence length $l_{\textrm{coh}}$. Thus we have
$\gamma (\Delta L)=\exp[-\frac{1}{2}\left(\Delta
L/l_{\textrm{coh}}^{p} \right)^{2}]$ and $\gamma ' (\Delta L')=
\exp[-\frac{1}{2}\left(\Delta L'/l_{\textrm{coh}} \right)^{2}]$.
The general form of Eq.~(\ref{coin-count-rate}) remains the same
even for other phase matching conditions. However, for other phase
matching conditions, the functional form of $\gamma ' (\Delta L')$
may not remain Gaussian or it may not even remain centered at
$\Delta L'=0$ \cite{shih-cw}.

We now look at the effects of varying $\Delta L$ and $\Delta L'$
on the coincidence count rate $R_{\textrm{AB}}$ of
Eq.~(\ref{coin-count-rate}) by considering two limiting cases.

\textit{Case} I. For $\Delta L'=0$ and  $\Delta\phi=0$,
\begin{eqnarray}\label{coin-count-rate2}
R_{\textrm{AB}}= C \ [ \ 1 + \gamma\left(\Delta L\right)
  \cos\left({k_{0}\Delta L}\right) \ ].
\end{eqnarray}
Interference is observed in the coincidence count rate as $\Delta
L$ is varied and gets washed out once $\Delta L$ exceeds the pump
coherence length. Thus $\Delta L$ plays the same role in
two-photon interference as does the optical path-length difference
in one-photon interference. It is because of this analogy that we
call $\Delta L$ the biphoton path-length difference. The
coincidence fringes seen in Franson-type
interferometers~\cite{Martienssen,kwiat-interference} and in the
double-pass setup \cite{frustrated two-photon} are examples of
effects due to variations in $\Delta L$.
\begin{figure}[b!]
\includegraphics{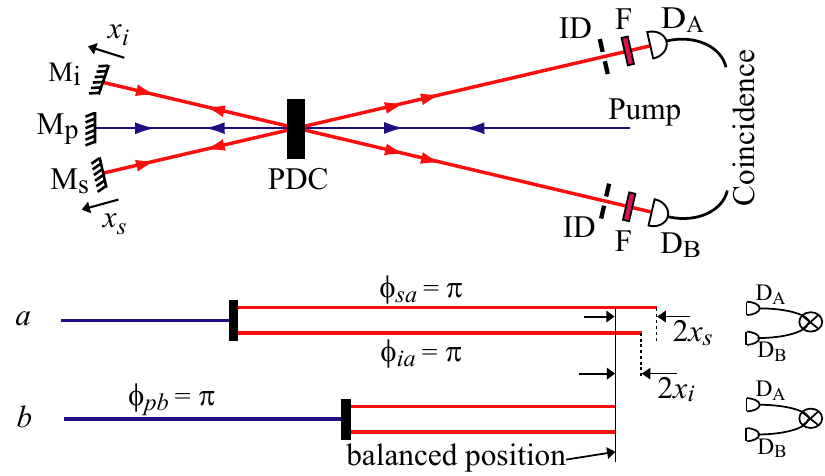}
  \caption{( Color online) Schematic of the experimental setup and the corresponding
  two-photon path
  diagrams. In alternative \textit{a}, the pump
photon gets down-converted in the forward pass while in
alternative \textit{b}, it gets down-converted in the backward
pass. F is an interference filter with 10 nm bandwidth, centered
at 727.6 nm; ID is an iris diaphragm.}\label{experimental-setup}
\end{figure}

\textit{Case} II. For $\Delta L$ and $\Delta\phi$ fixed,
\begin{eqnarray}\label{coin-count-rate3}
R_{\textrm{AB}} = C \ [ \  1+ K \gamma '\left(\Delta L'\right) \
],
\end{eqnarray}
where $K=\gamma (\Delta L)\cos{(k_{0}\Delta L+\Delta \phi)}$ is
constant. The coincidence count rate can show a dip when the two
alternatives interfere destructively ($K< 0$), and a hump when the
two alternatives interfere constructively ($K> 0$), as $\Delta L'$
is varied. These profiles, with widths equal to
$l_{\textrm{coh}}$, represent how the coherence between two
biphoton alternatives changes with a variation in $\Delta L'$.
$\Delta L'$ has no one-photon counterpart. Effects observed in the
HOM experiment ~\cite{HOM} and in the postponed compensation
experiment ~\cite{shih-can two-photon} are examples of two-photon
interference effects due to variations in $\Delta L'$. In the HOM
experiment ~\cite{HOM}, the effect was seen with $\Delta L=0$,
whereas in the postponed compensation experiment ~\cite{shih-can
two-photon}, the effects were seen in the limit
$l_{\textrm{coh}}\ll \Delta L\ll l_{\textrm{coh}}^{p}$.

In the HOM experiment the signal and idler photons from the PDC
are mixed at a beam splitter. Both photons, in the balanced
position of the setup, always leave through the same output port
of the beam splitter. As a result, a null is observed in the
coincidence count rate at the balanced position, leading to a dip
in the coincidence count rate as a function of the beam splitter
position. An intuitive explanation of this effect can be given in
terms of the bunching of signal and idler photons at a beam
splitter \cite{bunching}. However, the bunching interpretation is
not adequate for the postponed compensation  \cite{shih-can
two-photon} and related experiments \cite{shih-dip} in which
HOM-like effects are observed,%
\begin{widetext}

\begin{figure}[t!]
\includegraphics{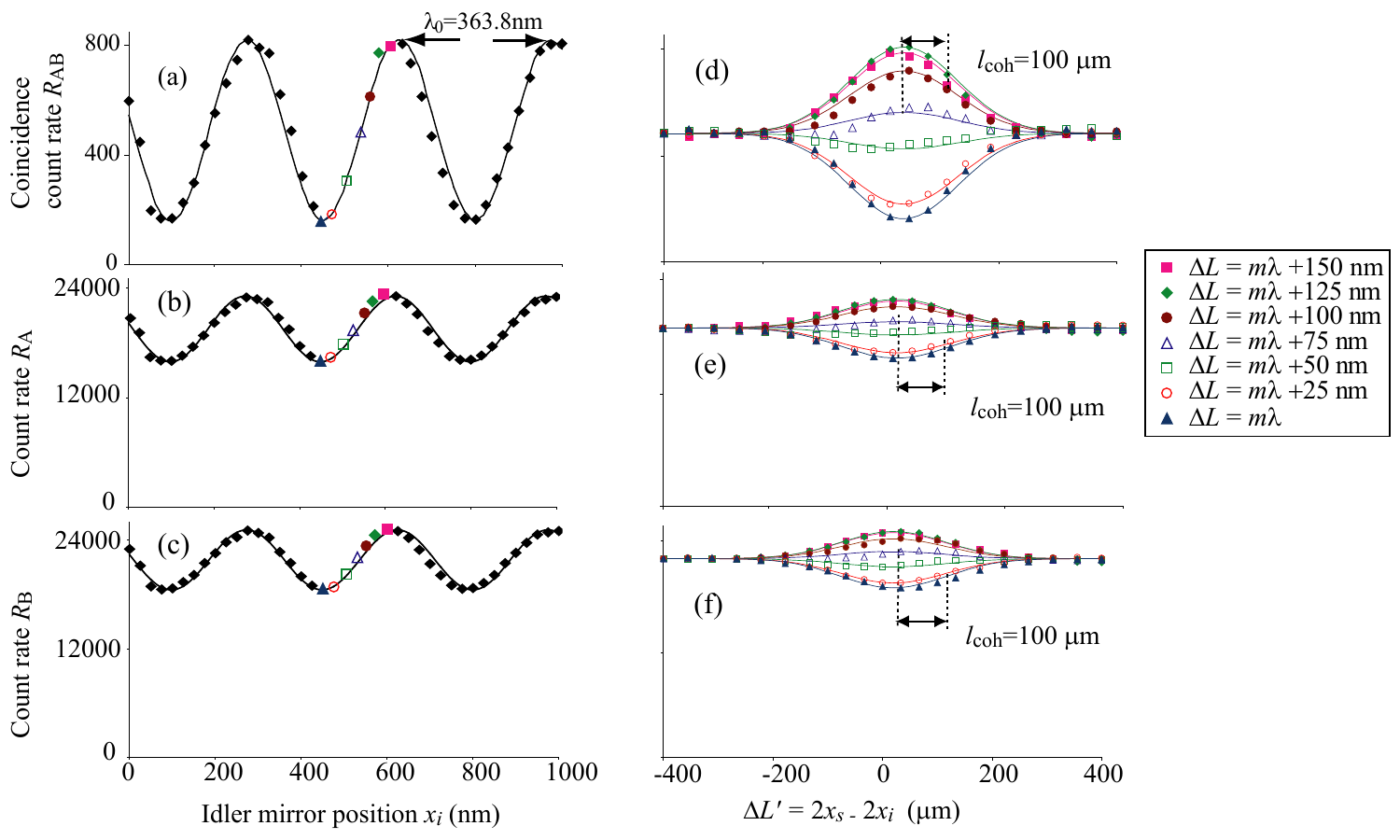}
  \caption{(Color online) Measured  (a) coincidence count rate $R_{\textrm{AB}}$, (b) count rate
  $R_{\textrm{A}}$,
   and (c) count rate $R_{\textrm{B}}$ as a function of the idler mirror position.
  Measured (d) coincidence count rate $R_{\textrm{AB}}$, (e) count rate $R_{\textrm{A}}$, and (f) count rate $R_{\textrm{B}}$,
  as a function of $\Delta
  L'$ for various fixed values of $\Delta L$. Solid lines are the theoretical best
  fits.
   }\label{experimental-result}
\end{figure}
\end{widetext}   even when signal and idler photons
do not simultaneously arrive  at a beam splitter. As discussed
above, both HOM and HOM-like effects are consequences of how
two-photon coherence changes as a function of the biphoton
path-asymmetry-length difference $\Delta L'$.

All these experiments required mixing of signal and idler photons
at a beam splitter. In contrast, we next report our experimental
observations of changes in two-photon coherence as a function of
$\Delta L'$, in a double-pass setup (shown in
Fig.~\ref{experimental-setup}), which does not involve mixing of
signal and idler photons at a beam splitter. A similar setup was
used earlier to demonstrate frustrated two-photon creation
\cite{frustrated two-photon}. In this setup, there are many ways
in which $\Delta L$ and $\Delta L'$ could be varied either
independently or simultaneously, by displacing the signal
($M_{s}$), idler ($M_{i}$), and pump ($M_{p}$) mirrors. In our
experiments we change only the signal and idler mirror positions.

In the balanced position of the setup of
Fig.~\ref{experimental-setup}, the distances of the signal, idler,
and pump mirrors from the crystal remain equal. We denote the
displacements of the signal and idler mirrors from the balanced
position by $x_{s}$ and $x_{i}$, respectively. Using the
two-photon path diagrams (lower part of
Fig.~\ref{experimental-setup}), we find that $\Delta
L=x_{s}+x_{i}$, $\Delta L'=2x_{s}-2x_{i}$, and $ \Delta \phi=\pi$.
Assuming $\Delta L$ to be always much smaller than the pump
coherence length $l_{\textrm{coh}}^{p}$, which in our experiment
is about 5 cm, we calculate the coincidence count rate
$R_{\textrm{AB}}$ for detectors $D_{\textrm{A}}$ and
$D_{\textrm{B}}$ using Eq.~(\ref{coin-count-rate}) as
\begin{equation}\label{frustrated-count-rate}
R_{\textrm{AB}}=C \{\ 1-
\gamma'(2x_{s}-2x_{i})\cos{[k_{0}(x_{s}+x_{i})]}\ \}.
\end{equation}
A displacement of either the signal or idler mirror changes both
$\Delta L$ and $\Delta L'$; therefore the coincidence count rate
$R_{\textrm{AB}}$ will show fringes as a function of the idler
mirror position. However, equal displacements of the signal and
idler mirrors in opposite directions change $\Delta L'$ while
keeping $\Delta L$ $(=x_{s}+x_{i})$ fixed; and therefore, the
coincidence count rate $R_{\textrm{AB}}$ will show either a dip or
a hump as a function of the joint displacement, depending on the
fixed value of $\Delta L$.

Figure ~\ref{experimental-result} shows the results of our
experimental investigations. A cw Ar-ion laser operating at
$\lambda_{0}=$ 363.8 nm was used as a pump to produce degenerate
type-I PDC. The signal and idler photons were collected into
multimode fibers and detected using two avalanche photodiodes. The
distance between the crystal and each detector was about 1.2 m.
With the diaphragms set to a size of about 1.2 mm, the effective
bandwidth of the signal-idler field becomes 0.85 nm, resulting in
a coherence length $l_{\textrm{coh}}$ of about 100 $\mu$m.

The idler mirror was first scanned around the balanced position
and as a result fringes were observed in the coincidence count
rate $R_{\textrm{AB}}$ [Fig.~\ref{experimental-result}(a)]. Next,
the idler mirror was placed at different fixed positions
corresponding to different values of $\Delta L$, as shown in
Fig.~\ref{experimental-result}(a) and in the inset, where $m$ is
an integer. Starting from each idler mirror position, the signal
and idler mirrors were displaced equally in opposite directions.
Dip and hump profiles of width $100$ $\mu$m were observed in the
coincidence count rate $R_{\textrm{AB}}$
[Fig.~\ref{experimental-result}(d)].

In addition, profiles similar to that of the coincidence count
rate were also observed in the single photon count rates
$R_{\textrm{A}}$ and $R_{\textrm{B}}$. As a function of the idler
mirror position, fringes were observed in the count rates of
$R_{\textrm{A}}$ [Fig.~\ref{experimental-result}(b)] and
$R_{\textrm{B}}$ [Fig.~\ref{experimental-result}(c)]; and as a
function of the simultaneous displacements of the signal and idler
mirrors, dip and hump profiles of width 100 $\mu$m were observed
in the count rates $R_{\textrm{A}}$
[Fig.~\ref{experimental-result}(e)] and $R_{\textrm{B}}$
[Fig.~\ref{experimental-result}(f)].

These one-photon effects cannot be described by second-order (in
the field) coherence theory \cite{wolf}, because the one-photon
path-length differences involved in these experiments are much
longer than the coherence lengths of the one-photon fields.
Interference effects in one-photon count rates have previously
also been observed in many two-photon experiments including
induced coherence \cite{induced coherence}, frustrated two-photon
creation \cite{frustrated two-photon}, and interference experiment
from separate pulses \cite{shih-single-photon}. Although these
one-photon effects have been interpreted differently, they can all
be described entirely in terms of the sum of two-photon
interference profiles. Thus, we represent the one-photon count
rate $R_{\textrm{X}}$ at a detection position $\textrm{X}$ as the
sum of the coincidence count rates $R_{\textrm{XY}_{i}}$ between
$\textrm{X}$ and all the other positions $\textrm{Y}_{i}$ to which
the twin of the photon detected at $\textrm{X}$ can go, i.e.,
\begin{equation}\label{single photon count rate}
    R_{\textrm{X}} = \sum_{i} R_{\textrm{XY}_{i}}.
\end{equation}
Summing over the detector positions $R_{\textrm{Y}_{i}}$ in
Eq.~(\ref{single photon count rate}) is the same as taking the
partial trace over all the possible modes of the twin. The
summation turns into an integral whenever the twin has finite
probabilities of arriving at a continuous set of detection points.

Now, for the setup in Fig.~\ref{experimental-setup}, the twin of a
photon detected at $D_{\textrm{A}}$ can go only to
$D_{\textrm{B}}$, and vice versa. Therefore, using
Eq.~(\ref{single photon count rate}) we find that the one-photon
count rates $R_{\textrm{A}}$ and $R_{\textrm{B}}$ are each equal
to the coincidence count rate $R_{\textrm{AB}}$. Hence, as a
function of either $\Delta L$ or $\Delta L'$, the one-photon count
rates show profiles similar to that of the coincidence count rate.

The dip-hump visibilities for $R_{\textrm{AB}}$, $R_{\textrm{A}}$
and, $R_{\textrm{B}}$ were found to be 67\%, 18\%, and 15\%,
respectively. The overall interference visibilities are low due to
imperfect mode matching of the fields in the two alternatives. The
visibilities of one-photon count rates are much smaller than that
of the coincidence count rate, because of the limited detection
efficiency of the system, which affects the one-photon count rate
more strongly than the coincidence count rate. The less than
perfect experimental fits are due to the uncontrollable drifts of
translation stages.

We see that both in this experiment and in the HOM \cite{HOM} and
HOM-like \cite{shih-can two-photon,shih-dip} experiments the same
effect, that is, the change in two-photon coherence as a function
of the biphoton path-asymmetry-length difference, is observed.
However, in our experiment---in contrast to the earlier
experiments---this effect is observed in a setup that does not
involve mixing of signal and idler photons at a beam splitter.
Moreover, to the best of our knowledge, we have observed for the
first time that the changes in two-photon coherence can manifest
themselves in the count rates of individual detectors as dip and
hump profiles.

In conclusion, we have constructed two length parameters to
describe two-photon interference involving the biphotons created
by PDC. We have shown that temporal two-photon coherence depends
only on these two parameters and that two-photon interference
effects, including one-photon interference effects observed in
certain two-photon experiments, can be fully characterized in
terms of these two parameters. We have performed an experiment in
which the variation of these two parameters could be independently
controlled and studied. We have further reported experimental
observations of HOM-like effects both in coincidence and in
one-photon count rates, and we have argued that HOM and HOM-like
effects can be best understood as observations of how two-photon
coherence changes with a variation in the biphoton
path-asymmetry-length difference.

We gratefully acknowledge financial support through a MURI grant
from the U.S. Army Research Office and through an STTR grant from
the U.S. Air Force Office of Scientific Research. We thank Mark
Gruneisen for useful discussions. K.W.C. acknowledges the support
of the Croucher Foundation.

\end{document}